\documentclass[12pt]{article}
\usepackage{amsmath}
\newcommand{\ket}[1]{\left| #1 \right\rangle}

\newcommand{\dket}[1]{\left.\left| #1 \right\rangle\right\rangle}

\def\nof{{\stackrel{\scriptscriptstyle{\circ}}
{\scriptscriptstyle{\circ}}}} 
\def\no{\raisebox{-2pt}{$\nof\:\!$}} 

\begin{document}

\begin{titlepage}

\begin{flushright}
TIT/HEP-507\\
{\tt hep-th/0309017}\\
September, 2003\\
\end{flushright}
\vskip 1cm

\begin{center}
{\Large \bf Closed String Emission from Unstable D-brane with Background Electric Field} \\
\vskip 2cm
{\large \bf Kenji~Nagami} \\
\vskip 0.5cm
{\large \it Department of Physics, Tokyo Institute of Technology,\\
Meguro, Tokyo 152-8551, Japan} \\
\vskip 2mm
{\small {\tt nagami@th.phys.titech.ac.jp}} \\
\end{center}

\vskip 1cm

\begin{abstract}
We study the closed string emission from an unstable D$p$-brane with constant background electric field in bosonic string theory.
The average total number density and the average total energy density of emitted closed strings are explicitly calculated in the presence of electric field.
It is explicitly shown that the energy density in the UV region becomes finite whenever the background electric field is switched on.
The energy density converted into closed strings in the presence of electric field is negligibly small compared with the D-brane tension in the weak string coupling limit.
\end{abstract}

\end{titlepage}

\normalsize
\newpage
\setcounter{page}{1}
%%%%%%%%%%%%%%%%%%%%%%%%%%%%%%%%%%%%%%%%%%%%%
%%%%%%%%%%%%%%%%%%%%%%%%%%%%%%%%%%%%%%%%%%%%%
%%%%%%%%%%%%%%%%%%%%%%%%%%%%%%%%%%%%%%%%%%%%%
\section{Introduction}
Tachyon condensation has been an issue in string theory~\cite{conjecture,Witten:1998cd}.
The decay process of an unstable D-brane can be described by tachyon condensation in the world volume effective field theory. It is observed that the decay leaves such objects called tachyon matter behind that they have the same energy density as the tension of the initial unstable D-brane, and localize on a hyper plane along which the initial unstable D-brane was extended, and have vanishing pressure~\cite{Sen:2002nu}--\cite{Minahan:2002if}.

For the case of the unstable D-brane with conserved electric flux on its world volume, it is shown that the fundamental string fluid also remains~\cite{Gibbons:2000hf}.
It is expected that the electric flux confines in spatially inhomogeneous tachyon condensation and behaves as fundamental string~\cite{Yi:1999hd}--\cite{Rey:2003zj}.
Then the decay of the unstable D-brane with conserved electric flux includes intriguing physics.

The time dependent decay process called rolling tachyon can also be described by a boundary CFT with time dependent boundary interaction~\cite{Sen:2002nu,Sen:2002in,Sen:2002vv}. Then the resulting boundary states have time dependent wave functions, which play roles as sources in equations of motion in closed string field theory. As is well known in field theory, time dependent sources radiate particles, then the radiation of closed strings would be certainly expected in string theory.
The closed string one point function on disk is of order $g_s^0$, then the radiation of closed strings is possible even in the weak string coupling limit $g_s \to 0$.

From thermodynamic point of view, the pair creation of open string massive modes is studied in the rolling tachyon background~\cite{Strominger:2002pc,Maloney:2003ck}.
The boundary CFT with the time dependent boundary interaction for half S-brane is identified with a timelike boundary Liouville theory that is obtained by a special analytic continuation of the ordinary spacelike one~\cite{Gutperle:2003xf,Strominger:2003fn,Schomerus:2003vv}.
Then the study on the rolling tachyon can be also performed by using the Liouville theory and a related $c=1$ matrix model~\cite{matrix}.

It would be significant to consider whether energy stored in the initial unstable D-brane is diffused into the transverse directions in the decay process.
Then it is necessary to evaluate the energy density carried by emitted closed string and compare it with the tension of the initial unstable D-brane.
The energy density is approximately calculated by using the boundary CFT with the time dependent boundary interaction, and turns out to have the UV divergence for $p\leq 2$.
The average transverse velocity of emitted closed string massive modes is also evaluated, and it is found that the emitted closed string massive modes slowly move apart from the initial unstable D-brane~\cite{Lambert:2003zr,Gaiotto:2003rm}.

In this paper we study the decay of the unstable D$p$-brane with constant background electric field on its world volume in bosonic string theory.
In section 2 we briefly review the analysis of the D$p$-brane decay using the boundary CFT, and the generalization to the case with the background electric field~\cite{Mukhopadhyay:2002en,Rey:2003xs}.
In section 3 we determine the average total number density and the average total energy density of emitted closed string modes.
It is explicitly shown that the part coming from the closed string UV region becomes finite for any values of $p$ whenever the background electric field is switched on.
The part coming from the closed string IR region is shown to be finite for $p<23$.
We also evaluate the average transverse velocity of emitted closed string massive modes in the background electric field, and observe that they still localize in high density near the hyper plane along which the initial unstable D-brane was extended.
We shall mainly consider the case of the half S-brane, and the result for the full S-brane is summarized in Appendix.
%%%%%%%%%%%%%%%%%%%%%%%%%%%%%%%%%%%%%%%%%%%%%
%%%%%%%%%%%%%%%%%%%%%%%%%%%%%%%%%%%%%%%%%%%%%
%%%%%%%%%%%%%%%%%%%%%%%%%%%%%%%%%%%%%%%%%%%%%
\section{Rolling tachyon boundary state}
We begin with the rolling tachyon on an unstable D$p$-brane in bosonic string theory without a background electric field.
The system can be described by a boundary CFT with a time dependent boundary interaction, which we briefly review.
We shall mainly study the particular case (called half S-brane)~\cite{Strominger:2002pc,Larsen:2002wc} in which the open string tachyon sitting at the unstable vacuum in the infinite past rolls down the potential to the locally stable vacuum in the infinite future, in a spatially homogeneous manner. This is described by a boundary CFT with an exactly marginal boundary interaction

\begin{equation}
S_{\rm bnd}=\lambda\int_{\partial\Sigma} d\tau \, e^{X^0}(\tau) \label{S-half},
\end{equation}
where we have the boundary of the worldsheet $\partial\Sigma$ parameterized by $\tau$. We use the $\alpha'=1$ unit throughout this paper. The overall parameter $\lambda$ might seem to specify an initial condition, but it can be absorbed into time translation, then it has no physical sense.
We can also study the case (called full S-brane)~\cite{Sen:2002nu,Sen:2002in} in which the open string tachyon sitting at a locally stable vacuum in the infinite past is lifted to the unstable vacuum by some incoming closed string radiation, then successively rolls down the potential with closed string radiation to the locally stable vacuum in the infinite future, in a spatially homogeneous manner~\cite{Gutperle:2002ai}. The result for the full S-brane is summarized in Appendix.
%%% boundary state without electric field %%%%%%%%%%%%
The boundary state for a D$p$-brane in this boundary CFT is represented as
\begin{equation}
\ket{B}= N_p \ket{B}^0 \otimes\Pi_{i=1}^p \ket{N}^i \otimes \Pi_{a=p+1}^{25} 
\ket{D}^a \otimes \ket{B}^{gh} \label{full-boundary-state},
\end{equation}
where $N_p$ is a normalization constant $N_p=\pi^{11/2}(2\pi)^{6-p}$~\cite{DiVecchia:1997pr}.
$\ket{N}$ and $\ket{D}$ denote the usual boundary states in free CFTs for directions with Neumann and Dirichlet boundary conditions, respectively. $\ket{B}^{gh}$ is in $bc$-ghosts CFT~\cite{Callan:1987px,Callan:1988wz}.

$\ket{B}^0$ is in the boundary CFT with the time dependent boundary interaction (\ref{S-half}). One can construct it by noting that the boundary interaction term becomes one of the generators of $SU(2)_1$ current algebra. In compact space with self-dual radius,  the boundary state is specified as the $SU(2)$ rotated Neumann state, then the form in non-compact space is obtained by projection of it.
That is specified as a linear combination of the Ishibashi states with coefficients of $SU(2)$ rotation matrix elements~\cite{Callan:1994ub,Polchinski:my},
\begin{equation}
\ket{B}^0=\sum_{j=0,\frac{1}{2},1,\cdots}\sum_{m=-j}^j 
D_{m,-m}^j (R) \dket{j;m,m}, \label{B-off-D}
\end{equation}
$\dket{j;m,m}$ denotes the Ishibashi state for Virasoro algebra~\cite{Ishibashi:1988kg} and $D_{m,-m}^j (R)$ denotes matrix element of a spin $j$ representation with a rotation matrix~\cite{encyclopedia,Recknagel:1998ih}
\begin{equation}
R=\left( \begin{array}{cc} 1 & 2\pi i\lambda \\
                             0 & 1
           \end{array}
    \right) \label{rotation-matrix}.
\end{equation}
One can decompose (\ref{B-off-D}) in terms of oscillators,
\begin{equation}
\ket{B}^0=\rho(\hat{x}^0)\ket{0,0}+\sigma(\hat{x}^0)\alpha_{-1}^0\bar{\alpha}_{-1}^0\ket{0,0}+\cdots , \label{B-off}
\end{equation}
where $\ket{0,0}$ denotes the $SL(2,C)$ vacuum and $\cdots$ represents states with higher excited level.
$\alpha_n^{\mu}$ and $\bar{\alpha}_n^{\mu}$ are understood to denote the oscillators of holomorphic and antiholomorphic parts of $X^{\mu}$, respectively.
The wave functions in coordinate space turn out to take the following form by analytic continuation,
\[ \rho(x^0)=\frac{1}{1+\hat{\lambda}e^{x^0}}, \hspace{1cm} \sigma(x^0)=2-\rho(x^0), \hspace{1.5cm} (\hat{\lambda}=2\pi\lambda) .\]
In the representation (\ref{B-off-D}), each descendant state is constructed on a primary state with discrete labels.
Boundary state may have a representation constructed on a primary state with continuous labels~\cite{Fateev:2000ik,Teschner:2000md},
\begin{equation}
\ket{B'}=\int \frac{dk_x^0}{2\pi}\> U(k_x^0) \dket{k_x^0}, \label{continuous-off}
\end{equation}
where $\dket{k_x^0}$ denotes a Ishibashi state constructed on a highest weight state $\ket{k_x^0}$. $k_x^0$ denotes a momentum eigenvalue for the direction $X^0$. Although the two representations of boundary state (\ref{B-off-D}) and (\ref{continuous-off}) are constructed on different vacuum representations of Virasoro algebra, they give a same value for physical observable as seen later.
The Virasoro operators are understood to be expanded in oscillators,
\begin{equation}
L_n^{x^0}=-\frac{1}{2}\sum_{m\in \rm{Z}} \no \alpha_{m+n}^0\alpha_{-m}^0 \no,
\hspace{0.5cm}
L_n^{x^1}=\frac{1}{2}\sum_{m\in \rm{Z}} \no \alpha_{m+n}^1\alpha_{-m}^1 \no,
\label{oscillator-expansion}
\end{equation}
where $\no$~$\no$ denotes the normal ordering.
The antiholomorphic parts are as well.
The coefficient $U(k_x^0)$ is to be determined, and related to one-point function of a primary field on disk,
\[U(k_x^0)^{\ast}=\left< B' \right.\left| k_x^0 \right>=\left<:e^{-ik_x^0X^0}(0,0):\right>_{\rm disk},\]
where $:$~$:$ denotes the normal ordered product.
By using the form of boundary state (\ref{B-off}), it is determined to be
\[U(k_x^0)=e^{-ik_x^0\ln \hat{\lambda}}\frac{\pi}{\sinh(\pi k_x^0)}.\]
The overall parameter $\lambda$ in (\ref{S-half}) appears only in the phase factor as expected, then it does not appear in the physical amplitude.
%%% boundary state with electric field %%%%%%%%%%%%
We shall study the rolling tachyon in a constant background electric field.
This system can be described by a boundary CFT with a following world sheet action
\begin{equation}
S=\frac{1}{4\pi}\int_{\Sigma} d^2\sigma\, \partial_a Y^{\mu}\partial_b Y^{\nu}(\delta^{ab}\eta_{\mu\nu}+i\epsilon^{ab}F_{\mu\nu})
+\lambda\int_{\partial\Sigma}d\tau\, e^{\sqrt{1-e^2}Y^0(\tau)},
\end{equation}
where we can use the Lorentz rotation symmetry to set the direction of the constant background electric field to be $Y^1$ without losing any generalities, that is, non-vanishing components are $F_{01}=-F_{10}=e$. $F_{\mu\nu}$ is understood to be scaled by $2\pi$ for simplicity. $\epsilon^{ab}$ is an usual antisymmetric tensor on world sheet, hence $\epsilon^{12}=-\epsilon^{21}=1$ on flat space. The boundary interaction term is still marginal.

The constant background electric field can be turned on by using T-duality and Lorentz boost\footnote{In the case of free CFT, the explicit form of the boundary state with background gauge field can be seen in, for instance,~\cite{Callan:1988wz,DiVecchia:1999uf,DiVecchia:1999fx}.}. Then the boundary state (\ref{full-boundary-state}) can be converted into that with the constant electric field by simple replacements~\cite{Rey:2003xs},
\begin{eqnarray}
\ket{0,0} \rightarrow \gamma^{-1}\ket{0,0}, &
\hat{x}^0 \rightarrow \gamma^{-1}\hat{y}^0, \nonumber \\
\left( \begin{array}{c} \alpha_n^0 \\ \alpha_n^1 \end{array} \right)\rightarrow \Lambda^{-1} \left( \begin{array}{c} \beta_n^0 \\ \beta_n^1 \end{array}\right),
& \hspace*{3mm}
\left( \begin{array}{c} \bar{\alpha}_n^0 \\ \bar{\alpha}_n^1 \end{array} \right)
\rightarrow \Lambda \left( \begin{array}{c} \bar{\beta}_n^0 \\ \bar{\beta}_n^1 \end{array}\right),
\label{replacement}
\end{eqnarray}
\begin{equation}
\Lambda=\gamma\left(\begin{array}{cc}1&e \\ e&1 \end{array}\right),
\hspace{5mm}
\Lambda^{-1}=\gamma\left(\begin{array}{cc}1&-e \\ -e&1 \end{array}\right),
\hspace{5mm}
\gamma^{-1}=\sqrt{1-e^2}, \nonumber
\end{equation}
where $\beta_n^{\mu}$ and $\bar{\beta}_n^{\mu}$ are understood to denote the oscillators of holomorphic and antiholomorphic parts of $Y^{\mu}$, respectively.
The resulting boundary state for the $Y^0$ and $Y^1$ directions can be decomposed in terms of oscillators,
\begin{align}
\ket{B}_e^{Y^0,Y^1}=&\{\rho_e(\hat{y}^0)+\sigma_e(\hat{y}^0)\gamma^2(\beta_{-1}^0-e\beta_{-1}^1)(\bar{\beta}_{-1}^0+e\bar{\beta}_{-1}^1)+\cdots \} \nonumber \\
&\times\{1-\gamma^2(\beta_{-1}^1-e\beta_{-1}^0)(\bar{\beta}_{-1}^1+e\bar{\beta}_{-1}^0)+\cdots \} \ket{0,0}, \label{B-on}
\end{align}
with modified wave functions,
\begin{equation}
\rho_e (y^0)=\gamma^{-1} \rho(\gamma^{-1}y^0),\hspace{0.5cm} \sigma_e(y^0)=\gamma^{-1}\sigma(\gamma^{-1}y^0). \label{modified-wave-fn}
\end{equation}
Note that the time evolution of wave functions is slowed down in the background electric field. This is caused by the fact that the effective tension of fundamental open string decreases with the strength of the background electric field, and vanishes in the critical limit, then the absolute value of mass square also has the same behavior~\cite{Nesterenko:pz}.

The prescription (\ref{replacement}) can also be used to construct a representation with continuous labels of boundary state for the $Y^0$ and $Y^1$ directions.
It turns out to be given by
\begin{equation}
\ket{B'}_e^{Y^0,Y^1}=\int \frac{dk_y^0\,dk_y^1}{(2\pi)^2}\> U(k_y^0,k_y^1) \dket{k_y^0,k_y^1}, \hspace{0.5cm}
U(k_y^0,k_y^1)=e^{-i\gamma k_y^0\ln \hat{\lambda}}\frac{2\pi^2}{\sinh(\pi\gamma k_y^0)}\delta(k_y^1),
\label{continuous-on}
\end{equation}
where the delta function on momentum comes from the translational invariance of the specified direction.
$\dket{k_y^0,k_y^1}$ is constructed from a direct product of Ishibashi states in the case without background electric field, $\dket{k_x^0}\otimes\dket{k_x^1}$, by using the prescription (\ref{replacement}) and replacements $k_x^0\to \gamma k_y^0$, $k_x^1\to k_y^1$ and $\hat{x}^1\to \hat{y}^1$ so that the highest weight state with specified momentum eigenvalues are properly normalized.
The acting Virasoro generators are such that each oscillator $\alpha$ is replaced with $\beta$ according to (\ref{replacement}) in the expanded forms (\ref{oscillator-expansion}), and let us denote them with superscript $'$.
The replacement (\ref{replacement}) may look like a Lorentz boost, but only with opposite boosted directions for holomorphic and antiholomorphic parts.
Then they satisfy the following relations,
\begin{equation}
L_n^{x^0{'}}+L_n^{x^1{'}}=L_n^{y^0}+L_n^{y^1},
\hspace{0.2cm}
[L_0^{y^0}+L_0^{y^1}, L_{-n}^{x^0{'}}]=nL_{-n}^{x^0{'}},
\hspace{0.2cm}
[L_0^{y^0}+L_0^{y^1}, L_{-n}^{x^1{'}}]=nL_{-n}^{x^1{'}},
\label{Lx-Ly}
\end{equation}
where $L_n^{y^0}$ and $L_n^{y^1}$ denote Virasoro generators for the $Y^0$ and $Y^1$ directions, and are understood to be expanded in oscillators,
\begin{equation}
L_n^{y^0}=-\frac{1}{2}\sum_{m\in \rm{Z}} \no \beta_{m+n}^0\beta_{-m}^0 \no,
\qquad L_n^{y^1}=\frac{1}{2}\sum_{m\in \rm{Z}} \no \beta_{m+n}^1\beta_{-m}^1\no . \label{oscillator-expansion-y}
\end{equation}
The antiholomorphic parts are as well.
%%%%%%%%%%%%%%%%%%%%%%%%%%%%%%%%%%%%%%%%%
%%%%%%%%%%%%%%%%%%%%%%%%%%%%%%%%%%%%%%%%%
%%%%%%%%%%%%%%%%%%%%%%%%%%%%%%%%%%%%%%%%%
\section{Closed string emission amplitude}
We will study the closed string emission from the unstable D$p$-brane on which open string tachyon rolls down the potential in the background electric field. We work out in the lowest perturbation level of string coupling where interactions between emitted closed strings are neglected, as in~\cite{Lambert:2003zr,Gaiotto:2003rm}.
The average total number of emitted closed string modes $\bar{N}$ may be determined as a Lorentzian cylinder amplitude 
by virtue of the optical theorem,
\begin{align}
\bar{N}&= {\rm Im} \biggl \langle B' \biggl | \frac{b_0^+c_0^-}{2(L_0+\bar{L}_0-i\epsilon)}\biggr | B' \biggr \rangle \label{cylinder} \\
&= {\rm Im} \int_0^{\infty}dt\, \biggl \langle B' \biggl | e^{-2t(L_0+\bar{L}_0-i\epsilon)} b_0^+c_0^- \biggr | B' \biggr \rangle \nonumber.
\end{align}
One may straightforwardly calculate this by using the relations (\ref{Lx-Ly}), but would encounter an ill-defined contribution from oscillators of timelike direction $Y^0$,
\begin{equation}
\int dk_y^0\,\frac{\pi}{2\sinh^2(\pi\gamma k_y^0)}
\,e^{t(k_y^0)^2}\,\eta\left(i\frac{2t}{\pi}\right)^{-1}. \label{ill-integral}
\end{equation}
Then we invoke the regularization procedure using $q$-gamma function to obtain a well-defined result~\cite{Karczmarek:2003xm}.
Let us define the character as
\begin{equation}
\chi_{\alpha}(q)=\eta(q)^{-1}\, q^{-\alpha^2}.
\end{equation}
The character can be expressed by the modular transform of it,
\begin{equation}
\chi_{\alpha}(\tilde{q})=\sqrt{2}\int_{-\infty}^{\infty}d\nu \cosh(4\pi\alpha\nu)\chi_{i\nu}(q),
\end{equation}
where $\tilde{q}$ is a modular transform of $q$, and vice versa. It is suitable for the current problem  to take as
\begin{equation}
q=e^{-\frac{\pi^2}{t}},\hspace{1cm} \tilde{q}=e^{-4t}.
\end{equation}
Then the equation (\ref{ill-integral}) is represented as the $b\rightarrow 1$ limit of
\begin{equation}
\frac{\pi}{\sqrt{2}}\int_{-\infty}^{\infty}dk_y^0\, \int_{-\infty}^{\infty}d\nu\, \left( \frac{\cosh(2\pi k_y^0\nu)}{\sinh(\pi k_y^0 b)\sinh(\pi k_y^0/b)} \right) \chi_{i\gamma\nu}(q).
\end{equation}
This has a double pole at $k_y^0=0$ that does not contribute to the integral, hence can be subtracted without changing value of integral~\cite{Teschner:2000md}
\begin{equation}
\frac{\pi}{\sqrt{2}}\int_{-\infty}^{\infty}dk_y^0\,\int_{-\infty}^{\infty}d\nu\,\left(\frac{\cosh(2\pi k_y^0\nu)}{\sinh(\pi k_y^0 b)\sinh(\pi k_y^0/b)}-\frac{1}{(\pi k_y^0)^2}\right)\chi_{i\gamma\nu}(q). \label{subtraction}
\end{equation}
This can be represented by using an integral representation of the $q$-gamma function $S_b(x)$~\cite{Zamolodchikov:1995aa,Fateev:2000ik,Teschner:2000md},
\begin{equation}
\ln S_b(x)= \int_0^{\infty}\frac{dt}{t}\,\left(\frac{\sinh((Q_b-2x)t)}{2\sinh(tb)\sinh(t/b)}-\frac{(Q_b/2-x)}{t}\right),
\hspace{1cm} Q_b=b+\frac{1}{b}.
\end{equation}
Then the equation (\ref{subtraction}) becomes
\begin{equation}
\sqrt{2}\int_{-\infty}^{\infty}d\nu\, \chi_{i\gamma\nu}(q)\> \partial_{\nu}\ln S_b \left(\frac{Q_b}{2}-\nu \right). \label{integral-nu}
\end{equation}
The $q$-gamma function $S_b(x)$ has zeroes at $x=Q_b+nb+m/b$ and simple poles at $x=-nb-m/b$ for non-negative integers $n$ and $m$.
According to~\cite{Karczmarek:2003xm}, the contour is allowed to include the zeroes that give rise to poles in the integrand, then the integral (\ref{integral-nu}) leads to imaginary parts given by
\begin{equation}
2^{3/2}\pi \, \eta(q)^{-1}\sum_{m,n=0}^{\infty}e^{-((n+\frac{1}{2})b+(m+\frac{1}{2})/b)^2\frac{\pi^2\gamma^2}{t}}\hspace{0.3cm}
\stackrel{b\to 1}{\longrightarrow}
\hspace{0.2cm}
2^{3/2}\pi \, \eta(q)^{-1}\sum_{n=1}^{\infty}n\, e^{-\frac{\pi^2\gamma^2}{t}n^2}.
\end{equation}
This result incorporates with the contributions from space-like directions and ghosts to give the average number of emitted closed string modes per unit volume,
\begin{align}
\bar{N}/V_p 
&=2^{-12-p}\pi^{12-\frac{3}{2}p} \sum_{n=1}^{\infty}n\int_0^{\infty}dt\, t^{-\frac{26-p}{2}}\,e^{-\frac{\pi^2}{t}\gamma^2n^2}\, \eta\left(i\frac{2t}{\pi}\right)^{-24} \nonumber \\
&=2^{-p}\pi^{-p/2} \sum_{n=1}^{\infty}n \int_0^{\infty} ds\, s^{-1-\frac{p}{2}}
\, e^{-s \gamma^2 n^2}\, \eta\left(\frac{is}{2\pi}\right)^{-24}, \label{N-bar}
\end{align}
where $s$ is a length of annulus with width $\pi$ and $2t$ is a length of cylinder with circumference $2\pi$, hence they are related by $s=\frac{\pi^2}{t}$. Note that the contribution from the background electric field appears as a factor of $\gamma$ in the exponent.
%%% oscillator representation %%%%%%%%%%
The equation (\ref{N-bar}) implies that the excited states for the timelike direction $Y^0$ do not contribute to the amplitude. This is an analogue of the no-ghost theorem~\cite{Hwang:1991an,Evans:1998qu} in the case with the background electric field. Then, regarding the closed string emission amplitude, it is adequate to take account of the time dependence coming only from states without any excitations of $Y^0$ direction, that is, $\rho_e(y^0)$.
This plays a role as a time dependent source in equations of motion that can be specified by decomposing the equation of motion in closed  string field theory into those for component fields,
\[ (Q_B+\bar{Q}_B)\ket{\Psi}=\ket{B} . \]
The average total number of emitted closed string modes $\bar{N}$ may also be represented as a simple generalization~\cite{Lambert:2003zr,Gaiotto:2003rm} of the case in field theory,
\begin{equation}
\bar{N} = \sum_s \frac{1}{2E_s}  |\tilde{\rho}_e (E_s)|^2, \label{number}
\end{equation}
where $\tilde{\rho}_e(E)$ is Fourier transform of the time dependent source $\rho_e(y^0)$,
\begin{equation}
\tilde{\rho}_e(E) =\int dy^0 \rho_e(y^0) e^{iEy^0}
     =-i\, e^{-i\gamma E \ln \hat{\lambda}} \frac{\pi}{\sinh (\pi\gamma E)}, 
     \nonumber
\end{equation}
and $E$ is energy of emitted closed string mode with the same level $N$ for left and right movers and transverse momenta $k_\bot$,
\[ E^2=|k_\bot|^2+4(N-1). \]
The sum on $s$ in (\ref{number}) denotes integrals of transverse momenta and level sum with density of states $d_N$ defined by $\Pi_{n=1}^{\infty}(1-q^n)^{-24}=\sum_{N=0}^{\infty}d_Nq^N$. Each state excited in spatial directions is unit normalized, then the whole state with the same level $N$ contributes by $d_N$.
$V_p$ denotes a spatial volume of D$p$-brane, as $(2\pi)^p\delta^p(k_{\parallel}=0)=\int d^px_{\parallel}=V_p$, where the delta functions making tangential momenta zero come from translational invariance of Neumann directions. Fourier transform of the wave function  for Dirichlet directions is $e^{-ik_{\perp}x_0}$ if D-brane is located at $x_0$ in transverse directions, then they contribute by $1$ to the square of absolute value.
One can find $\bar{N}$ to be specified by the same equations as (\ref{N-bar}).
%%% energy %%%%%%%%%%%%%%%%%%
The average total energy of emitted closed string modes $\bar{E}$ is represented as well
\begin{equation}
\bar{E} = \frac{1}{2} \sum_s |\tilde{\rho}_e (E_s)|^2, \label{energy}
\end{equation}
and the similar calculation leads to
\begin{align}
\bar{E}/V_p
&=2^{-12-p}\pi^{13-\frac{3}{2}p}\sum_{n=1}^{\infty}n^2\int_0^{\infty}dt\, t^{-\frac{28-p}{2}}\gamma e^{-\frac{\pi^2}{t}\gamma^2n^2} \eta\left(i\frac{2t}{\pi}\right)^{-24} \label{E-bar-t} \\
&=2^{-p}\pi^{-1-\frac{p}{2}}\sum_{n=1}^{\infty}n^2\int_0^{\infty}ds\, s^{-\frac{p}{2}}\gamma e^{-s\gamma^2n^2} \eta\left(i\frac{s}{2\pi}\right)^{-24}. \label{E-bar-s}
\end{align}
We shall analyze the alternative regions of modulus $s(t)$ to study the convergence of $\bar{E}$, up to an overall numerical factor.
We explicitly study only the case of $\bar{E}$, but $\bar{N}$ is obtained by replacing $p$ with $p+2$ in $\bar{E}$, up to an overall numerical factor.
%%% s--->00 %%%%%%%%%%%%%
\begin{flushleft}
\underline{UV region: $s>\Lambda$ }
\end{flushleft}
\begin{equation}
\bar{E}/V_p 
\to \gamma \sum_{n=1}^{\infty}n^2\sum_{N=0}^{\infty}d_N\int_{\Lambda}^{\infty}ds\,s^{-\frac{p}{2}}e^{-s(\gamma^2n^2+N-1)} \label{E_UV},
\end{equation}
where a cut off parameter $\Lambda>0$ is introduced into the $s$-integral (\ref{E-bar-s}) in order to make the power series expansion well-defined. The sign $\to$ denotes that the right hand side does not include the contribution from the closed string IR region.
It follows from the equation (\ref{E_UV}) that there is no UV divergence under the $s$-integral whenever the background electric field is turned on, that is, $\gamma >1$. Note that in the case with no electric field ($\gamma=1$), the $s$-integral for $n=1$ and $N=0$ gives rise to a closed string UV divergence for $p\leq 2$, and this fact corresponds to the result obtained in~\cite{Lambert:2003zr,Gaiotto:2003rm}.
The right hand side of (\ref{E_UV}) can be represented as
\begin{equation}
\gamma\sum_{n=1}^{\infty}n^2\sum_{N=0}^{\infty}d_N\,(\gamma^2n^2+N-1)^{\frac{p-2}{2}}\,\Gamma\left(\frac{2-p}{2},\Lambda(\gamma^2n^2+N-1)\right),
\label{infinite_sum}
\end{equation}
by using the incomplete Gamma function defined by
\begin{equation}
\Gamma(z,\alpha)=\int_{\alpha}^{\infty}ds\, s^{z-1}e^{-s},
\hspace{5mm} (\alpha >0).
\end{equation}
We shall aim to show the convergence of the infinite sum (\ref{infinite_sum}),
then it is adequate to show the convergence of a partial sum in which $n$ and $N$ run from some finite numbers $n'$ and $N'$ to $\infty$.
With large $n'$ and $N'$, the asymptotic forms of $\Gamma(z,\alpha)$ and $d_N$ are available in the infinite sum. The partial sum that might give rise to divergence is represented as
\begin{equation}
\gamma\Lambda^{-\frac{p}{2}}\sum_{n=n'}^{\infty}\sum_{N=N'}^{\infty}\frac{n^2N^{-\frac{27}{4}}}{\gamma^2n^2+N}\,e^{4\pi\sqrt{N}}e^{-\Lambda(\gamma^2n^2+N)/2}. \label{asymptotic_sum}
\end{equation}
This is smaller than an infinite sum given by
\begin{equation}
\gamma\Lambda^{-\frac{p}{2}}\sum_{n=n'}^{\infty}\sum_{N=N'}^{\infty} n^2N^{-\frac{27}{4}}\,e^{4\pi\sqrt{N}}e^{-\Lambda(\gamma^2n^2+N)/2}. \label{double_sum}
\end{equation}
This can be represented as a double sum made of a positive term series on $n$ and one on $N$ that are convergent, respectively. Then the double sum (\ref{double_sum}) is convergent, hence the right hand side of (\ref{E_UV}) is also convergent.

Accordingly, it appears that the energy density coming from the closed string UV region is finite whenever the background electric field is turned on.
The original unstable D-brane tension is infinite in the week string coupling limit $g_s\to 0$. Then it follows that the only a negligible portion of the energy density stored in the unstable D$p$-brane is converted into the closed string UV region for any values of $p$.
%%%%% transverse momentum square %%%%%%%%%%%%%%%%%%%%%
One can evaluate the transverse velocity of closed string massive modes for $p<25$. The production probability of closed string modes with transverse momentum $k$ and level $N$ is found from the equation (\ref{number}) to be proportional to
\begin{equation}
\left[\sqrt{k^2+4(N-1)}\,\sinh^2(\pi\gamma\sqrt{k^2+4(N-1)})\right]^{-1}=:P_N(k),\end{equation}
then the expectation value of transverse momentum square can be defined as
\begin{equation}
\left\langle k^2 \right\rangle_N=\frac{\int d^{25-p}k \; k^2 P_N(k)}{\int d^{25-p}k \, P_N(k)}. \label{k2}
\end{equation}
For a fixed high level $N$, $P_N(k)$ is approximately given by
\begin{equation}
P_N(k) \sim \frac{4}{\sqrt{k^2+4N}}\,e^{-2\pi\gamma\sqrt{k^2+4N}}.
\end{equation}
By using this formula for $P_N(k)$, the denominator of (\ref{k2}) is specified as
\begin{equation}
c\gamma^{-\frac{24-p}{2}}N^{\frac{24-p}{4}}\; K_{(24-p)/2}(4\pi\gamma\sqrt{N})
\approx c' N^{\frac{23-p}{4}}\gamma^{-\frac{25-p}{2}}e^{-4\pi\gamma\sqrt{N}},
\end{equation}
where $c$ and $c'$ are numerical constants, $\approx$ means the saddle point estimation, and $K_{\nu}(z)$ denotes the modified Bessel function defined by
\begin{equation}
K_{\nu}(z)=2^{-\nu-1}z^{\nu}\int_0^{\infty}dt\,t^{-\nu-1}\exp \left(-\frac{z^2}{4t}-t\right).
\end{equation}
The numerator of (\ref{k2}) is specified as well,
\begin{eqnarray}
&& \left[\frac{1}{(2\pi)^2}\frac{d^2}{d\gamma^2}-4N \right] \int d^{25-p}k\, P_N(k) \\
&\approx&\! c' \left[\frac{25-p}{\pi} N^{\frac{25-p}{4}} \gamma^{-\frac{27-p}{2}}+ \frac{(25-p)(27-p)}{16\pi^2} N^{\frac{23-p}{4}}\gamma^{-\frac{29-p}{2}}   \right ] e^{-4\pi\gamma\sqrt{N}}.
\end{eqnarray}
Then the expectation value of transverse momentum square behaves up to the leading term for a fixed high level $N$ as
\begin{equation}
\langle\, k^2 \rangle_N \sim
\frac{25-p}{\pi}\,N^{1/2}\gamma^{-1}
+\frac{(25-p)(27-p)}{16\pi^2}\gamma^{-2} \hspace{0.5cm} (N \gg 1).
\end{equation}
For nonrelativistic particles, the corresponding velocity is determined by simply dividing momentum by mass ($\sim N^{1/2}$), then the resulting expectation value of transverse velocity for closed string massive modes behaves as $\gamma^{-1/2}N^{-1/4}$. This result corresponds with~\cite{Lambert:2003zr,Gaiotto:2003rm} in the limit of vanishing background electric field ($e\to 0$, hence $\gamma\to 1$). The emitted closed string massive modes would move apart from the unstable D-brane slowly and slowly with mass, then the collection of them has high density even in the background electric field.
%%% s-->0 %%%%%%%%%%%%%%%
\begin{flushleft}
\underline{IR region: $t>\Lambda'$ }
\end{flushleft}
\begin{equation}
\bar{E}/V_p \to \gamma \sum_{n=1}^{\infty}n^2 \sum_{N=0}^{\infty}d_N \int_{\Lambda'}^{\infty}dt\, t^{-\frac{28-p}{2}}\exp\left[-\frac{\pi^2\gamma^2n^2}{t}-4(N-1)t \right], \label{E_IR}
\end{equation}
where a cut off $\Lambda'>0$ is introduced in the $t$-integral (\ref{E-bar-t}) in order to make the power series expansion well-defined. The sign $\to$ denotes that the right hand side does not include the contribution coming from the closed string UV region.
The $N=0$ part comes from the closed string tachyon, and it diverges.
The closed string tachyon does not appear in superstring theory, then we would not suffer from the divergence from it.
The $N=1$ part comes from the closed string massless mode, and it can be represented as
\begin{equation}
\gamma^{p-25} \sum_{n=1}^{\infty}n^{p-24}\;\gamma\left(\frac{26-p}{2},\frac{\pi^2\gamma^2n^2}{\Lambda'}\right), \label{N1}
\end{equation}
by using the incomplete gamma function defined by
\begin{equation}
\gamma(z,\alpha)=\int_0^{\alpha}ds\, s^{z-1}e^{-s}=\Gamma(z)-\Gamma(z,\alpha),
\hspace{5mm} (\alpha >0).
\end{equation}
We shall aim to study the convergence of the infinite sum (\ref{N1}), then it is adequate to study the convergence of a partial sum in which $n$ runs from a finite number $n'$ to $\infty$.
With large $n'$, the asymptotic form of $\gamma(z,\alpha)$ is available in the infinite sum. Then the partial sum that might give rise to divergence is represented as
\begin{equation}
\gamma^{p-25}\Gamma\left(\frac{26-p}{2}\right)\sum_{n=n'}^{\infty}n^{p-24}
-\pi^{24-p}\gamma^{-1}{\Lambda'}^{\frac{p-24}{2}}\sum_{n=n'}^{\infty}e^{-\frac{\pi^2\gamma^2n^2}{2\Lambda'}}.
\end{equation}
The second sum is convergent, while the first is convergent only for $p<23$.
The apparent divergence for $p\geq 23$ certainly means that the back reaction is to be involved and the naive assumption of flat spacetime is not consistent with D-brane of codimension less than $3$.
Incidentally, the radiation of closed string massless states from the full S-brane type of rolling tachyon boundary state was examined in~\cite{Chen:2002fp}, where the energy density loss was shown to be finite for $p<23$, with which our result is consistent.

Next, let us estimate the energy density of closed string massive modes. It is given by a partial sum in which $N$ runs from $2$ to $\infty$,
\begin{equation}
\gamma \sum_{n=1}^{\infty} n^2 \sum_{N=2}^{\infty} d_N \,(N-1)^{\frac{26-p}{2}} \int_{4(N-1)\Lambda'}^{\infty} dt \, t^{-\frac{28-p}{2}} \exp \left[-\frac{4\pi^2 \gamma^2 n^2 (N-1)}{t}-t \right]. \label{massive_IR}
\end{equation}
This is smaller than an infinite sum that is defined by taking a limit $\Lambda'\to 0$ in (\ref{massive_IR}), which is represented by using the modified Bessel function,
\begin{equation}
\gamma^{-\frac{24-p}{2}}\sum_{n=1}^{\infty}n^{\frac{p-22}{2}}\sum_{N=2}^{\infty}d_N\, (N-1)^{\frac{26-p}{4}}\,K_{(26-p)/2}\left(4\pi\gamma n \sqrt{N-1}\right). \label{IR_infinite_sum}
\end{equation}
We shall aim to study the convergence of the infinite sum (\ref{massive_IR}), it is adequate to study the convergence of a partial sum of (\ref{IR_infinite_sum}) in which $n$ and $N$ run from some finite numbers $n'$ and $N'$ to $\infty$.
With large $n'$ and $N'$, the asymptotic forms of $K_{\nu}(z)$ and $d_N$ are available in the infinite sum. The part in the infinite sum (\ref{IR_infinite_sum}) that might give rise to divergence is represented by
\begin{equation}
\gamma^{-\frac{25-p}{2}}\sum_{n=n'}^{\infty}\sum_{N=N'}^{\infty} n^{-\frac{23-p}{2}}N^{-\frac{2+p}{4}} e^{-4\pi\gamma n\sqrt{N}}. \label{IR_infinite_sum2}
\end{equation}
This is smaller than an infinite sum given by
\begin{equation}
\gamma^{-\frac{25-p}{2}}\sum_{n=n'}^{\infty}\sum_{N=N'}^{\infty} n^{-\frac{23-p}{2}}N^{-\frac{2+p}{4}} e^{-4\pi\gamma n}e^{-4\pi\gamma\sqrt{N}}.\label{IR_infinite_sum3}
\end{equation}
This can be represented as a double sum made of a positive term series on $n$ and one on $N$ that are convergent respectively, then this double sum (\ref{IR_infinite_sum3}) is convergent.
This fact explicitly shows that the energy density coming from the closed string massive modes in the IR region (\ref{massive_IR}) is finite even if the infinite sum on $n$ and level $N$ is involved.

The original $\bar{E}/V_p$ (\ref{E-bar-t}) or (\ref{E-bar-s}) can be represented as a sum of the right hand sides of (\ref{E_UV}) and (\ref{E_IR}) with $\Lambda'=\pi^2/\Lambda$ and suitable numerical coefficients.
The energy density converted into the closed strings in the presence of electric flux is explicitly shown to be negligibly small compared with the D-brane tension which is infinite in the weak string coupling limit.
Incidentally, it can be seen that the energy density decreases with the strength of the background electric field, and vanishes in the critical limit. This behavior comes from the slowly time dependence of source in the background electric field (\ref{modified-wave-fn}).

%%% B+- %%%%%%%%%%%
The form of (\ref{N-bar}) implies that the average total number can be equated with the following amplitude in free CFT, in a similar way to~\cite{Lambert:2003zr,Gaiotto:2003rm},
\begin{equation}
\bar{N}= \biggl \langle B_- \biggl | \frac{b_0^+c_0^-}{2(L_0+\bar{L}_0)} \biggr | B_+ \biggr \rangle, \label{D-array}
\end{equation}
where boundary states $\ket{B_{\pm}}$ are defined by
\[ \ket{B_{\pm}}=\sum_{n=0}^{\infty}\ket{B_p[x^0=\pm i(2n+1)\pi\gamma] \, } ,\]
which describe a semi-infinite array of S$(p-1)$-branes\footnote{S$(p-1)$-brane has $p$ spatial directions with Neumann condition and $26-p$ directions including time with Dirichlet condition, as originally defined in \cite{Gutperle:2002ai}.} sitting along imaginary time axis with period $2\pi\gamma$. We emphasize that the period becomes longer in the background electric field.
The amplitude (\ref{D-array}) can be interpreted as a sum of 1-loop amplitudes of open strings stretching between each pair of S$(p-1)$-branes at positive and negative imaginary time.
The IR dominant contribution comes from the lightest open string stretching between the two closest S$(p-1)$-branes at $x^0=\pm i\pi\gamma$. The mass square of this lightest open string mode is lifted to a positive finite value $\gamma^2-1$, then no IR divergence appears in this open string 1-loop amplitude. This corresponds to the fact that the closed string UV divergence of $\bar{N}$ and $\bar{E}$ disappears whenever the background electric field is turned on.

%%%%%%%%%%%%%%%%%%%%%%%%%%%%%%%%%%%%%%
%%%%%%%%%%%%%%%%%%%%%%%%%%%%%%%%%%%%%%
%%%%%%%%%%%%%%%%%%%%%%%%%%%%%%%%%%%%%%
\section{Conclusion}
We have studied the tachyon condensation in the presence of background electric field on an unstable D$p$-brane in bosonic string theory. The average total number density and the average total energy density of emitted closed strings were explicitly calculated.
It was explicitly shown that the energy density in the UV region became finite whenever the background electric field was turned on.
The UV finiteness in the presence of electric field was also expressed in the associated open string 1-loop amplitude.
Incidentally, the energy density in the IR region was also explicitly shown to be finite for $p<23$, though the apparent divergence for $p=23, 24$ certainly indicated that the naive assumption of flat spacetime was inconsistent with D-brane of codimension less than $3$.

The energy density converted into closed strings in the presence of electric flux turned out to be negligibly small compared to the D-brane tension that was infinite in the week string coupling limit.
The electric flux is generically identified with the conserved fundamental string charge to which the Kalb-Ramond field couples.
Accordingly, it appears that the apparently missing part of the D-brane tension would be exhausted to generate stretched fundamental strings under tachyon condensation.
This observation is consistent with the analysis in~\cite{Sen:2003xs}, where the direction parallel to the electric flux was taken to be compact, and most part of the initial D-brane tension was observed to be converted into winding states of closed string, which carried fundamental string charges with finite energy costed.
We note that the spacetime was noncompact in our analysis, then no winding state was involved in the boundary state with which we calculated the emitted energy density.

It was recently proposed that the collection of emitted closed string with very high density could be effectively described with classical open string theories, and this description was also valid in the presence of electric field~\cite{Sen:2003bc}.
We explicitly showed that the average transverse velocity of emitted closed string massive modes was sufficiently small even in the electric field, so that emitted massive closed strings were collected with very high density.
%%%%%%%%%%%%%%%%%%%%%%%%%%%%%%%%%%%%%%
%%%%%%%%%%%%%%%%%%%%%%%%%%%%%%%%%%%%%%
%%%%%%%%%%%%%%%%%%%%%%%%%%%%%%%%%%%%%%
\section*{Acknowledgements}
\indent

The author would like to thank to K. Ito and K. Hotta for useful comments.

%%%%%%%%%%%%%%%%%%%%%%%%%%%%%%%%%%%%%%%%%
%%%%%%%%%%%%%%%%%%%%%%%%%%%%%%%%%%%%%%%%%
%%%%%%%%%%%%%%%%%%%%%%%%%%%%%%%%%%%%%%%%%
\section*{Appendix: Full S-brane}
So far we consider the particular case (called half S-brane) of tachyon condensation in which the open string tachyon sitting at the unstable vacuum in the infinite past rolls down the potential to the locally stable vacuum in the infinite future, in a spatially homogeneous manner.
We can also consider the case (called full S-brane) in which the open string tachyon sitting at the locally stable vacuum in the infinite past is lifted to the unstable vacuum by some incoming closed string radiation, then successively rolls down the potential with closed string radiation to the locally stable vacuum in the infinite future, in a spatially homogeneous manner.
We summarize the result for the full S-brane.
This system is described by a boundary CFT with a boundary interaction~\cite{Sen:2002nu,Sen:2002in} instead of (\ref{S-half})
\[ S_{\rm bnd}=\tilde{\lambda}\int_{\partial\Sigma} d\tau \, \cosh X^0(\tau). \]
The boundary state for a D$p$-brane in this theory takes a similar form to (\ref{full-boundary-state}), but with the rotation matrix $R$ replaced with
\[R=\left( \begin{array}{cc} \cos (\pi\tilde{\lambda}) & i\sin (\pi\tilde{\lambda}) \\
                             i\sin (\pi\tilde{\lambda}) & \cos (\pi\tilde{\lambda})
           \end{array}
    \right) . \]
The wave functions are given by
\[ \rho(x^0)=\frac{1}{1+\hat{\lambda}e^{x^0}}+\frac{1}{1+\hat{\lambda}e^{-x^0}}-1, \hspace{0.5cm} \sigma(x^0)=1+\cos(2\pi\tilde{\lambda})-\rho(x^0), \hspace{1cm} (\hat{\lambda}=\sin(\pi\tilde{\lambda}) ,\]
and those in the background electric field are specified as in (\ref{modified-wave-fn}).
The average total number $\bar{N}$ and the average total energy $\bar{E}$ of closed string can be determined in the formulae (\ref{number}) and (\ref{energy}).
$\bar{N}$ is given by
\begin{equation}
\bar{N}/V_p
=2^{-p}\pi^{-p/2} \sum_{n=1}^{\infty}n \int_0^{\infty} ds\, s^{-1-\frac{p}{2}}  \,\eta\left(\frac{is}{2\pi}\right)^{-24}
\left( 2e^{-n^2\gamma^2s}-e^{-(n+\frac{i}{\pi}\ln\hat{\lambda})^2\gamma^2s}-e^{-(n-\frac{i}{\pi}\ln\hat{\lambda})^2\gamma^2s} \right).
\end{equation}
$\bar{E}$ is given by
\begin{align}
\bar{E}/V_p
&=2^{-p}\pi^{-1-\frac{p}{2}}\sum_{n=1}^{\infty} \int_0^{\infty}ds\, s^{-\frac{p}{2}}\,\gamma\, \eta\left(i\frac{s}{2\pi}\right)^{-24} \nonumber \\
&\times\left(
2n^2e^{-s\gamma^2n^2}-n(n+a)e^{-s\gamma^2(n+a)^2}-n(n-a)e^{-s\gamma^2(n-a)^2}
\right), \hspace{0.5cm} a=\frac{i}{\pi}\ln\hat{\lambda}.
\end{align}
$\bar{E}$ and $\bar{N}$ are zero in the case of $\hat{\lambda}=1$, and may have divergence from massive closed strings in the case of smaller $\hat{\lambda}$.
%%%%%%%%%%%%%%%%%%%%%%%%%%%%%%%%%%%%%%
%%%%%%%%%%%%%%%%%%%%%%%%%%%%%%%%%%%%%%
%%%%%%%%%%%%%%%%%%%%%%%%%%%%%%%%%%%%%%
\setlength{\baselineskip}{14pt}


\begin{thebibliography}{99}
\bibitem{conjecture}
%\cite{Sen:1998sm}
%\item
A.~Sen,
``Tachyon condensation on the brane antibrane system,''
JHEP {\bf 9808}, 012 (1998)
[arXiv:hep-th/9805170];
%%CITATION = HEP-TH 9805170;%%
%\cite{Sen:1999mh}
%\item
% A.~Sen,
``Descent relations among bosonic D-branes,''
Int.\ J.\ Mod.\ Phys.\ A {\bf 14}, 4061 (1999)
[arXiv:hep-th/9902105];
%%CITATION = HEP-TH 9902105;%%
%\cite{Sen:1999mg}
%\item
% A.~Sen,
``Non-BPS states and branes in string theory,''
arXiv:hep-th/9904207.
%%CITATION = HEP-TH 9904207;%%


%\cite{Witten:1998cd}
\bibitem{Witten:1998cd}
E.~Witten,
``D-branes and K-theory,''
JHEP {\bf 9812}, 019 (1998)
[arXiv:hep-th/9810188].
%%CITATION = HEP-TH 9810188;%%


%\cite{Sen:2002nu}
\bibitem{Sen:2002nu}
A.~Sen,
``Rolling tachyon,''
JHEP {\bf 0204}, 048 (2002)
[arXiv:hep-th/0203211].
%%CITATION = HEP-TH 0203211;%%

%\cite{Sen:2002in}
\bibitem{Sen:2002in}
A.~Sen,
``Tachyon matter,''
JHEP {\bf 0207}, 065 (2002)
[arXiv:hep-th/0203265].
%%CITATION = HEP-TH 0203265;%%

%\cite{Sen:2002an}
\bibitem{Sen:2002an}
A.~Sen,
``Field theory of tachyon matter,''
Mod.\ Phys.\ Lett.\ A {\bf 17}, 1797 (2002)
[arXiv:hep-th/0204143].
%%CITATION = HEP-TH 0204143;%%

%\cite{Ohta:2002ac}
\bibitem{Ohta:2002ac}
K.~Ohta and T.~Yokono,
``Gravitational approach to tachyon matter,''
Phys.\ Rev.\ D {\bf 66}, 125009 (2002)
[arXiv:hep-th/0207004].
%%CITATION = HEP-TH 0207004;%%


%\cite{Ishida:2002fr}
\bibitem{Ishida:2002fr}
A.~Ishida and S.~Uehara,
``Gauge fields on tachyon matter,''
Phys.\ Lett.\ B {\bf 544}, 353 (2002)
[arXiv:hep-th/0206102].
%%CITATION = HEP-TH 0206102;%%

%\cite{Ishida:2003cj}
\bibitem{Ishida:2003cj}
A.~Ishida and S.~Uehara,
``Rolling down to D-brane and tachyon matter,''
JHEP {\bf 0302}, 050 (2003)
[arXiv:hep-th/0301179].
%%CITATION = HEP-TH 0301179;%%

%\cite{Sugimoto:2002fp}
\bibitem{Sugimoto:2002fp}
S.~Sugimoto and S.~Terashima,
``Tachyon matter in boundary string field theory,''
JHEP {\bf 0207}, 025 (2002)
[arXiv:hep-th/0205085].
%%CITATION = HEP-TH 0205085;%%

%\cite{Minahan:2002if}
\bibitem{Minahan:2002if}
J.~A.~Minahan,
``Rolling the tachyon in super BSFT,''
JHEP {\bf 0207}, 030 (2002)
[arXiv:hep-th/0205098].
%%CITATION = HEP-TH 0205098;%%


%\cite{Gibbons:2000hf}
\bibitem{Gibbons:2000hf}
G.~W.~Gibbons, K.~Hori and P.~Yi,
``String fluid from unstable D-branes,''
Nucl.\ Phys.\ B {\bf 596}, 136 (2001)
[arXiv:hep-th/0009061].
%%CITATION = HEP-TH 0009061;%%


%\cite{Yi:1999hd}
\bibitem{Yi:1999hd}
P.~Yi,
``Membranes from five-branes and fundamental strings from Dp branes,''
Nucl.\ Phys.\ B {\bf 550}, 214 (1999)
[arXiv:hep-th/9901159].
%%CITATION = HEP-TH 9901159;%%


%\cite{Bergman:2000xf}
\bibitem{Bergman:2000xf}
O.~Bergman, K.~Hori and P.~Yi,
``Confinement on the brane,''
Nucl.\ Phys.\ B {\bf 580}, 289 (2000)
[arXiv:hep-th/0002223].
%%CITATION = HEP-TH 0002223;%%


%\cite{Harvey:2000jt}
\bibitem{Harvey:2000jt}
J.~A.~Harvey, P.~Kraus, F.~Larsen and E.~J.~Martinec,
``D-branes and strings as non-commutative solitons,''
JHEP {\bf 0007}, 042 (2000)
[arXiv:hep-th/0005031].
%%CITATION = HEP-TH 0005031;%%


%\cite{Larsen:2000sq}
\bibitem{Larsen:2000sq}
F.~Larsen,
``Fundamental strings as noncommutative solitons,''
Int.\ J.\ Mod.\ Phys.\ A {\bf 16}, 650 (2001)
[arXiv:hep-th/0010181].
%%CITATION = HEP-TH 0010181;%%


%\cite{Sen:2000kd}
\bibitem{Sen:2000kd}
A.~Sen,
``Fundamental strings in open string theory at the tachyonic vacuum,''
J.\ Math.\ Phys.\  {\bf 42}, 2844 (2001)
[arXiv:hep-th/0010240].
%%CITATION = HEP-TH 0010240;%%


%\cite{Gibbons:2002tv}
\bibitem{Gibbons:2002tv}
G.~Gibbons, K.~Hashimoto and P.~Yi,
``Tachyon condensates, Carrollian contraction of Lorentz group, and fundamental strings,''
JHEP {\bf 0209}, 061 (2002)
[arXiv:hep-th/0209034].
%%CITATION = HEP-TH 0209034;%%


%\cite{Hashimoto:2002sk}
\bibitem{Hashimoto:2002sk}
K.~Hashimoto, P.~M.~Ho and J.~E.~Wang,
``S-brane actions,''
Phys.\ Rev.\ Lett.\  {\bf 90}, 141601 (2003)
[arXiv:hep-th/0211090].
%%CITATION = HEP-TH 0211090;%%


%\cite{Rey:2003zj}
\bibitem{Rey:2003zj}
S.~J.~Rey and S.~Sugimoto,
``Rolling of modulated tachyon with gauge flux and emergent fundamental  string,''
Phys.\ Rev.\ D {\bf 68}, 026003 (2003)
[arXiv:hep-th/0303133].
%%CITATION = HEP-TH 0303133;%%


%\cite{Sen:2002vv}
\bibitem{Sen:2002vv}
A.~Sen,
``Time evolution in open string theory,''
JHEP {\bf 0210}, 003 (2002)
[arXiv:hep-th/0207105].
%%CITATION = HEP-TH 0207105;%%


%\cite{Strominger:2002pc}
\bibitem{Strominger:2002pc}
A.~Strominger,
``Open string creation by S-branes,''
arXiv:hep-th/0209090.
%%CITATION = HEP-TH 0209090;%%


%\cite{Maloney:2003ck}
\bibitem{Maloney:2003ck}
A.~Maloney, A.~Strominger and X.~Yin,
``S-brane thermodynamics,''
JHEP {\bf 0310}, 048 (2003)
[arXiv:hep-th/0302146].
%%CITATION = HEP-TH 0302146;%%


%\cite{Gutperle:2003xf}
\bibitem{Gutperle:2003xf}
M.~Gutperle and A.~Strominger,
``Timelike boundary Liouville theory,''
Phys.\ Rev.\ D {\bf 67}, 126002 (2003)
[arXiv:hep-th/0301038].
%%CITATION = HEP-TH 0301038;%%


%\cite{Strominger:2003fn}
\bibitem{Strominger:2003fn}
A.~Strominger and T.~Takayanagi,
``Correlators in timelike bulk Liouville theory,''
Adv.\ Theor.\ Math.\ Phys.\  {\bf 7}, 369 (2003)
[arXiv:hep-th/0303221].
%%CITATION = HEP-TH 0303221;%%


%\cite{Schomerus:2003vv}
\bibitem{Schomerus:2003vv}
V.~Schomerus,
``Rolling tachyons from Liouville theory,''
JHEP {\bf 0311}, 043 (2003)
[arXiv:hep-th/0306026].
%%CITATION = HEP-TH 0306026;%%


\bibitem{matrix}
%\cite{McGreevy:2003kb}
%\bibitem{McGreevy:2003kb}
J.~McGreevy and H.~Verlinde,
``Strings from tachyons: The c = 1 matrix reloated,''
arXiv:hep-th/0304224.
%%CITATION = HEP-TH 0304224;%%

%\cite{Klebanov:2003km}
%\bibitem{Klebanov:2003km}
I.~R.~Klebanov, J.~Maldacena and N.~Seiberg,
``D-brane decay in two-dimensional string theory,''
JHEP {\bf 0307}, 045 (2003)
[arXiv:hep-th/0305159].
%%CITATION = HEP-TH 0305159;%%

%\cite{McGreevy:2003ep}
%\bibitem{McGreevy:2003ep}
J.~McGreevy, J.~Teschner and H.~Verlinde,
``Classical and quantum D-branes in 2D string theory,''
arXiv:hep-th/0305194.
%%CITATION = HEP-TH 0305194;%%

%\cite{Constable:2003rc}
%\bibitem{Constable:2003rc}
N.~R.~Constable and F.~Larsen,
``The rolling tachyon as a matrix model,''
JHEP {\bf 0306}, 017 (2003)
[arXiv:hep-th/0305177].
%%CITATION = HEP-TH 0305177;%%

%\cite{Takayanagi:2003sm}
%\bibitem{Takayanagi:2003sm}
T.~Takayanagi and N.~Toumbas,
``A matrix model dual of type 0B string theory in two dimensions,''
JHEP {\bf 0307}, 064 (2003)
[arXiv:hep-th/0307083].
%%CITATION = HEP-TH 0307083;%%

%\cite{Douglas:2003up}
%\bibitem{Douglas:2003up}
M.~R.~Douglas, I.~R.~Klebanov, D.~Kutasov, J.~Maldacena, E.~Martinec and N.~Seiberg,
``A new hat for the c = 1 matrix model,''
arXiv:hep-th/0307195.
%%CITATION = HEP-TH 0307195;%%

%\cite{Gaiotto:2003yf}
%\bibitem{Gaiotto:2003yf}
D.~Gaiotto, N.~Itzhaki and L.~Rastelli,
``On the BCFT description of holes in the c = 1 matrix model,''
Phys.\ Lett.\ B {\bf 575}, 111 (2003)
[arXiv:hep-th/0307221].
%%CITATION = HEP-TH 0307221;%%


%\cite{Gutperle:2003ij}
%\bibitem{Gutperle:2003ij}
M.~Gutperle and P.~Kraus,
``D-brane dynamics in the c = 1 matrix model,''
arXiv:hep-th/0308047.
%%CITATION = HEP-TH 0308047;%%

%\cite{Sen:2003iv}
%\bibitem{Sen:2003iv}
A.~Sen,
``Open-Closed Duality: Lessons from Matrix Model,''
arXiv:hep-th/0308068.
%%CITATION = HEP-TH 0308068;%%

%\cite{Teschner:2003qk}
%\bibitem{Teschner:2003qk}
J.~Teschner,
``On boundary perturbations in Liouville theory and brane dynamics in noncritical string theories,''
arXiv:hep-th/0308140.
%%CITATION = HEP-TH 0308140;%%

%\cite{Okuyama:2003jk}
%\bibitem{Okuyama:2003jk}
K.~Okuyama,
``Comments on half S-branes,''
JHEP {\bf 0309}, 053 (2003)
[arXiv:hep-th/0308172].
%%CITATION = HEP-TH 0308172;%%


%\cite{Lambert:2003zr}
\bibitem{Lambert:2003zr}
N.~Lambert, H.~Liu and J.~Maldacena,
``Closed strings from decaying D-branes,''
arXiv:hep-th/0303139.
%%CITATION = HEP-TH 0303139;%%

%\cite{Gaiotto:2003rm}
\bibitem{Gaiotto:2003rm}
D.~Gaiotto, N.~Itzhaki and L.~Rastelli,
``Closed strings as imaginary D-branes,''
arXiv:hep-th/0304192.
%%CITATION = HEP-TH 0304192;%%

%\cite{Mukhopadhyay:2002en}
\bibitem{Mukhopadhyay:2002en}
P.~Mukhopadhyay and A.~Sen,
``Decay of unstable D-branes with electric field,''
JHEP {\bf 0211}, 047 (2002)
[arXiv:hep-th/0208142].
%%CITATION = HEP-TH 0208142;%%


%\cite{Rey:2003xs}
\bibitem{Rey:2003xs}
S.~J.~Rey and S.~Sugimoto,
``Rolling tachyon with electric and magnetic fields: T-duality approach,''
Phys.\ Rev.\ D {\bf 67}, 086008 (2003)
[arXiv:hep-th/0301049].
%%CITATION = HEP-TH 0301049;%%


%\cite{Larsen:2002wc}
\bibitem{Larsen:2002wc}
F.~Larsen, A.~Naqvi and S.~Terashima,
``Rolling tachyons and decaying branes,''
JHEP {\bf 0302}, 039 (2003)
[arXiv:hep-th/0212248].
%%CITATION = HEP-TH 0212248;%%


%\cite{Gutperle:2002ai}
\bibitem{Gutperle:2002ai}
M.~Gutperle and A.~Strominger,
``Spacelike branes,''
JHEP {\bf 0204}, 018 (2002)
[arXiv:hep-th/0202210].
%%CITATION = HEP-TH 0202210;%%


%\cite{DiVecchia:1997pr}
\bibitem{DiVecchia:1997pr}
P.~Di Vecchia, M.~Frau, I.~Pesando, S.~Sciuto, A.~Lerda and R.~Russo,
``Classical p-branes from boundary state,''
Nucl.\ Phys.\ B {\bf 507}, 259 (1997)
[arXiv:hep-th/9707068].
%%CITATION = HEP-TH 9707068;%%


%\cite{Callan:1987px}
\bibitem{Callan:1987px}
C.~G.~Callan, C.~Lovelace, C.~R.~Nappi and S.~A.~Yost,
``Adding Holes And Crosscaps To The Superstring,''
Nucl.\ Phys.\ B {\bf 293}, 83 (1987).
%%CITATION = NUPHA,B293,83;%%


%\cite{Callan:1988wz}
\bibitem{Callan:1988wz}
C.~G.~Callan, C.~Lovelace, C.~R.~Nappi and S.~A.~Yost,
``Loop Corrections To Superstring Equations Of Motion,''
Nucl.\ Phys.\ B {\bf 308}, 221 (1988).
%%CITATION = NUPHA,B308,221;%%


%\cite{Callan:1994ub}
\bibitem{Callan:1994ub}
C.~G.~Callan, I.~R.~Klebanov, A.~W.~Ludwig and J.~M.~Maldacena,
``Exact solution of a boundary conformal field theory,''
Nucl.\ Phys.\ B {\bf 422}, 417 (1994)
[arXiv:hep-th/9402113].
%%CITATION = HEP-TH 9402113;%%

%\cite{Polchinski:my}
\bibitem{Polchinski:my}
J.~Polchinski and L.~Thorlacius,
``Free Fermion Representation Of A Boundary Conformal Field Theory,''
Phys.\ Rev.\ D {\bf 50}, 622 (1994)
[arXiv:hep-th/9404008].
%%CITATION = HEP-TH 9404008;%%


%\cite{Ishibashi:1988kg}
\bibitem{Ishibashi:1988kg}
N.~Ishibashi,
``The Boundary And Crosscap States In Conformal Field Theories,''
Mod.\ Phys.\ Lett.\ A {\bf 4}, 251 (1989).
%%CITATION = MPLAE,A4,251;%%


\bibitem{encyclopedia}
L.~C.~Biedenharn and J.~D.~Louck,
``Encyclopedia of Mathematics and its Applications vol.8,''
Addison-Wesley Publishing Company, 1981


%\cite{Recknagel:1998ih}
\bibitem{Recknagel:1998ih}
A.~Recknagel and V.~Schomerus,
``Boundary deformation theory and moduli spaces of D-branes,''
Nucl.\ Phys.\ B {\bf 545}, 233 (1999)
[arXiv:hep-th/9811237].
%%CITATION = HEP-TH 9811237;%%


%\cite{Fateev:2000ik}
\bibitem{Fateev:2000ik}
V.~Fateev, A.~B.~Zamolodchikov and A.~B.~Zamolodchikov,
``Boundary Liouville field theory. I: Boundary state and boundary  two-point function,''
arXiv:hep-th/0001012.
%%CITATION = HEP-TH 0001012;%%


%\cite{Teschner:2000md}
\bibitem{Teschner:2000md}
J.~Teschner,
``Remarks on Liouville theory with boundary,''
arXiv:hep-th/0009138.
%%CITATION = HEP-TH 0009138;%%


%\cite{DiVecchia:1999uf}
\bibitem{DiVecchia:1999uf}
P.~Di Vecchia, M.~Frau, A.~Lerda and A.~Liccardo,
``(F,Dp) bound states from the boundary state,''
Nucl.\ Phys.\ B {\bf 565}, 397 (2000)
[arXiv:hep-th/9906214].
%%CITATION = HEP-TH 9906214;%%


%\cite{DiVecchia:1999fx}
\bibitem{DiVecchia:1999fx}
P.~Di Vecchia and A.~Liccardo,
``D-branes in string theory. II,''
arXiv:hep-th/9912275.
%%CITATION = HEP-TH 9912275;%%


%\cite{Nesterenko:pz}
\bibitem{Nesterenko:pz}
V.~V.~Nesterenko,
``The Dynamics Of Open Strings In A Background Electromagnetic Field,''
Int.\ J.\ Mod.\ Phys.\ A {\bf 4}, 2627 (1989).
%%CITATION = IMPAE,A4,2627;%%


%\cite{Karczmarek:2003xm}
\bibitem{Karczmarek:2003xm}
J.~L.~Karczmarek, H.~Liu, J.~Maldacena and A.~Strominger,
``UV finite brane decay,''
JHEP {\bf 0311}, 042 (2003)
[arXiv:hep-th/0306132].
%%CITATION = HEP-TH 0306132;%%


%\cite{Zamolodchikov:1995aa}
\bibitem{Zamolodchikov:1995aa}
A.~B.~Zamolodchikov and A.~B.~Zamolodchikov,
``Structure constants and conformal bootstrap in Liouville field theory,''
Nucl.\ Phys.\ B {\bf 477}, 577 (1996)
[arXiv:hep-th/9506136].
%%CITATION = HEP-TH 9506136;%%


%\cite{Hwang:1991an}
\bibitem{Hwang:1991an}
S.~Hwang,
``Cosets as gauge slices in SU(1,1) strings,''
Phys.\ Lett.\ B {\bf 276}, 451 (1992)
[arXiv:hep-th/9110039].
%%CITATION = HEP-TH 9110039;%%

%\cite{Evans:1998qu}
\bibitem{Evans:1998qu}
J.~M.~Evans, M.~R.~Gaberdiel and M.~J.~Perry,
``The no-ghost theorem for AdS(3) and the stringy exclusion principle,''
Nucl.\ Phys.\ B {\bf 535}, 152 (1998)
[arXiv:hep-th/9806024].
%%CITATION = HEP-TH 9806024;%%


%\cite{Chen:2002fp}
\bibitem{Chen:2002fp}
B.~Chen, M.~Li and F.~L.~Lin,
``Gravitational radiation of rolling tachyon,''
JHEP {\bf 0211}, 050 (2002)
[arXiv:hep-th/0209222].
%%CITATION = HEP-TH 0209222;%%


%\cite{Sen:2003xs}
\bibitem{Sen:2003xs}
A.~Sen,
``Open-closed duality at tree level,''
Phys.\ Rev.\ Lett.\  {\bf 91}, 181601 (2003)
[arXiv:hep-th/0306137].
%%CITATION = HEP-TH 0306137;%%


%\cite{Sen:2003bc}
\bibitem{Sen:2003bc}
A.~Sen,
``Open and closed strings from unstable D-branes,''
Phys.\ Rev.\ D {\bf 68}, 106003 (2003)
[arXiv:hep-th/0305011].
%%CITATION = HEP-TH 0305011;%%


%\cite{Moeller:2002vx}
\bibitem{Moeller:2002vx}
N.~Moeller and B.~Zwiebach,
``Dynamics with infinitely many time derivatives and rolling tachyons,''
JHEP {\bf 0210}, 034 (2002)
[arXiv:hep-th/0207107].
%%CITATION = HEP-TH 0207107;%%


%\cite{Okuda:2002yd}
\bibitem{Okuda:2002yd}
T.~Okuda and S.~Sugimoto,
``Coupling of rolling tachyon to closed strings,''
Nucl.\ Phys.\ B {\bf 647}, 101 (2002)
[arXiv:hep-th/0208196].
%%CITATION = HEP-TH 0208196;%%


%\cite{Kluson:2003rd}
\bibitem{Kluson:2003rd}
J.~Kluson,
``Particle production on half S-brane,''
arXiv:hep-th/0306002.
%%CITATION = HEP-TH 0306002;%%


%\cite{Kluson:2003sh}
\bibitem{Kluson:2003sh}
J.~Kluson,
``The Schroedinger wave functional and S-branes,''
Class.\ Quant.\ Grav.\  {\bf 20}, 4285 (2003)
[arXiv:hep-th/0307079].
%%CITATION = HEP-TH 0307079;%%


%\cite{Ohmori:2003je}
\bibitem{Ohmori:2003je}
K.~Ohmori,
``Toward open-closed string theoretical description of rolling tachyon,''
arXiv:hep-th/0306096.
%%CITATION = HEP-TH 0306096;%%


%\cite{Sugawara:2003xt}
\bibitem{Sugawara:2003xt}
Y.~Sugawara,
``Thermal partition functions for S-branes,''
JHEP {\bf 0308}, 008 (2003)
[arXiv:hep-th/0307034].
%%CITATION = HEP-TH 0307034;%%


%\cite{Hashimoto:2003qx}
\bibitem{Hashimoto:2003qx}
K.~Hashimoto, P.~M.~Ho, S.~Nagaoka and J.~E.~Wang,
``Time evolution via S-branes,''
Phys.\ Rev.\ D {\bf 68}, 026007 (2003)
[arXiv:hep-th/0303172].
%%CITATION = HEP-TH 0303172;%%


%\cite{Okuyama:2003wm}
\bibitem{Okuyama:2003wm}
K.~Okuyama,
``Wess-Zumino term in tachyon effective action,''
JHEP {\bf 0305}, 005 (2003)
[arXiv:hep-th/0304108].
%%CITATION = HEP-TH 0304108;%%

\end{thebibliography}
\end{document}